\documentclass[twocolumn,amsmath,amssymb,letterpaper,showpacs,prl]{revtex4}
\usepackage[tight]{subfigure}
\usepackage[]{graphicx}
\usepackage{bm}
\usepackage{dcolumn}

\begin{document}


\title{Consequences of Anomalous Diffusion in Disordered Systems Under Cyclic Forcing}
\author{Mitch Mailman}
\author{Michelle Girvan}
\author{Wolfgang Losert}
\email{wlosert@umd.edu}
\affiliation{Department of Physics, IREAP, and IPST, University of
Maryland, College Park, Maryland, 20742}
\date{\today}

\begin{abstract}
We use numerical simulations to study the behavior of 2D frictionless disk systems under cyclic shear as a function of reversal amplitude $\gamma_r$.  Our studies focus on mean bulk and disk dynamics.  These measurements suggest a crossover from a subdiffusive, $\gamma_r$ dependent regime to a regime where the grain motions are diffusive, with properties dependent only on total shear strain.  We discuss model stochastic processes that are consistent with these observations.  Finally, we introduce a modified Mean-Squared Displacement (mMSD) which takes into account the motion of the neighborhood of nearby grains and yields new insights into local displacement fluctuations.  We find that scaling properties of the displacement distributions are consistent with well studied stochastic models of anomalous diffusion and suggest scale-invariant cage dynamics.
\end{abstract}

\maketitle

{\it Introduction ---} In recent years there has been a good deal of interest in amorphous systems under oscillatory driving\cite{Behringer,GollubAndPine,SuspSims,CoreyCyclicSim,OurShearPaper,DauchotCyclicShear}.  These works focus on the property of reversibility of particle dynamics, or the dependence of dynamical behaviors on the distance to $\phi_J$, which is the density at which the packing becomes ``jammed."  However, what has been less well studied is the systematic dependence of bulk and particle dynamics of amorphous systems on driving amplitude, for systems that are not strictly reversible.  In this study, we simulate an amorphous system under slow, cyclic shear for a range of reversal amplitudes $\gamma_r$ (the strain at which shear is reversed).  We find that for our simple model, a crossover exists in the behavior of bulk and particle dynamics from a regime that is dependent on $\gamma_r$ to one that is not.  This crossover appears to influence a wide range of properties, including shear stress, packing fraction, and particle displacements.  
	
	Finally, we address the microscopic origins of the $\gamma_r$ dependent diffusive properties. We find a $\gamma_r$ dependence in the exponent $\beta$ of the mean-squared displacement (MSD) $\left<r^2\right>\propto t^\beta$.  To limit the possible underlying stochastic processes that can potentially explain the observed sub-diffusivity, we analyze the the scaling properties of probability distribution functions (PDFs) as a function of distance and time for all grains.  We observe that the exponent $\beta$ exclusively dictates the $\gamma_r$ dependence of the PDFs, as well as the MSDs and the dependence of bulk properties on $\gamma_r$.  

{\it Methods ---} We study a minimal model system with the use of molecular dynamics (MD) simulation.  Bidisperse (1:1) frictionless disks, with larger disks having 1.4 times larger diameter than the smaller, interact via a linear spring potential and damping\cite{DurianBubbles} that is dependent on the difference in velocity between contacting grains.  Lees-Edwards boundary conditions are used to apply a shear strain of $\gamma$.  It is important to note that we choose not to include a coupling to an affine displacement field since we only intend to capture interactions between grains at contact, and not interactions with a constituent fluid.  Packings are kept at approximately fixed pressure during shear by uniformly modulating the grain sizes in response to changes in pressure.  At each simulation time step, the change in pressure is calculated, and if the pressure decreases, then grain diameters are increased, otherwise the grain diameters are decreased (the pressure is kept approximately fixed to the same value throughout this work).     
	Under these conditions, we study 8000 grain systems over a range of $\gamma_r=\frac{\Delta x_r}{L}$ from 0.05 to 0.225.  

\begin{figure}[t]
\centering
\subfigure{\includegraphics[width=0.32\textwidth]{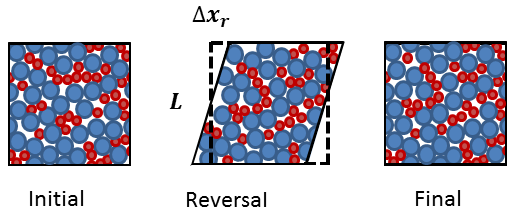}}
\vspace{-.3cm}
\caption{The schematic diagram illustrates the shear as applied to the 2D model granular system.  The initial state has a repeat cell which is square with linear size $L$.  The strain is applied by shifting the the lower and upper adjacent repeat cells horizontally by up to $\Delta x_r$ as per Lees-Edwards boundary conditions.  Then, the shear strain $\Delta x_r$ is reversed back to zero. }
\vspace{-.35cm}
\label{fig:Schematic}
\end{figure}	

\begin{figure}[t]
\centering
\subfigure{\includegraphics[width=0.4\textwidth]{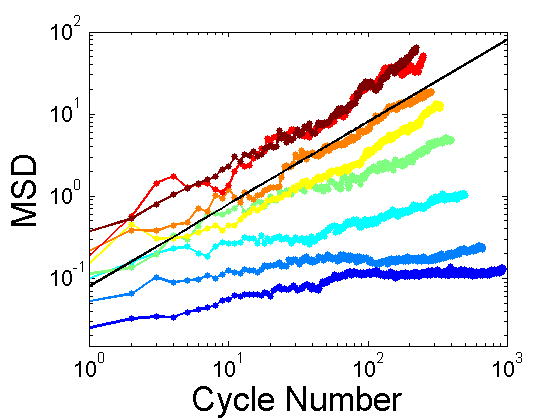}}\vspace{-.3cm}
\subfigure{\includegraphics[width=0.4\textwidth]{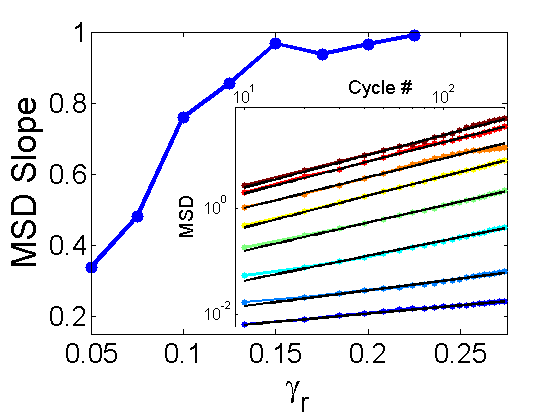}}
\vspace{-.5cm}
\caption{(Top) MSD measured relative to the initial, isotropic state.  Different colors correspond to different reversal amplitudes, with blue being the lowest and red being the highest.  Solid black lines with a slope of 1 are added for reference.  (Bottom,inset) Time (or cycle) averaged MSDs.  The best linear fits are shown for each reversal amplitude.  (Bottom, main) the slopes from the linear fits are shown as a function of reversal amplitude.  }
\vspace{-.35cm}
\label{fig:Dynamics}
\end{figure}

{\it Results ---} We begin by presenting some of the dynamical properties of our system.  In particular, we look at the mean squared displacements (MSD) of the grains.  We will always report MSDs on a cycle-by-cycle basis, that is to say, the MSD is calculated at the end of each cycle, where for reversible dynamics the system should return to zero net displacement.  We find that for all $\gamma_r$ studied grains are at least slightly displaced relative to the beginning of the cycle, with increasing MSDs as a function of cycle number (where cycle number replaces time).  First, we simply report the MSD relative to the initial (isotropic) state.  These MSDs are noisy, making it difficult to extract a single $\beta$ characterizing the evolution of the grain displacements.  As a result, we also average over intervals of up to 200 cycles, each separated by 10 cycles in the oscillatory shear trajectory, and produce a time (or cycle) averaged MSD (see figure \ref{fig:Dynamics}).  These averaged MSD curves exhibit a single exponent over the range of 200 cycles.  Neglecting the first 10 cycles as the very short time dynamics appear to be somewhat distinct from the longer time scale dynamics, we apply linear fits to the MSD curves and extract slopes.       
	These slopes as a function of $\gamma_r$ provide the first evidence of a crossover between two regimes as $\gamma_r$ is increased.  For very small $\gamma_r$, the dynamics are highly subdiffusive.  However, for a $\gamma_r$=0.15 the exponent $\beta$ is nearly 1, suggesting diffusive dynamics.  As $\gamma_r$ continues to increase, $\beta$ stays unchanged at approximately 1.  

\begin{figure}[t]
\centering
\subfigure{\includegraphics[width=0.40\textwidth]{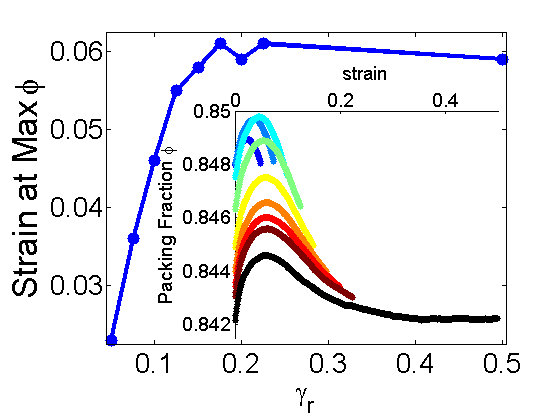}}\vspace{-.3cm}
\subfigure{\includegraphics[width=0.40\textwidth]{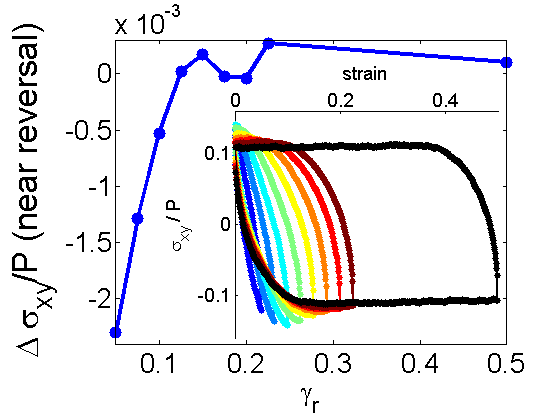}}
\vspace{-.5cm}
\caption{Packing fraction (top,inset) and shear stress $\sigma_{xy}$ (bottom,inset) averaged over many cycles as a function of strain $\gamma$, for varying $\gamma_r$.  Colors range from blue (smallest $\gamma_r$) to red (largest $\gamma_r$).  (Main) $\gamma_r$ dependence for strain (top) at $\phi$ and slope of $\sigma_{xy}$ (bottom) at reversal.}
\label{fig:DensityAndStress}
\vspace{-.25cm}
\end{figure}

	This crossover in grain dynamics as $\gamma_r$ is increased also manifests itself in the bulk properties of the shear flow (figure \ref{fig:DensityAndStress}).  During each cycle, the packing fraction $\phi$ increases to a peak value, which depends on $\gamma_r$, and then decreases again.  Averaged over many cycles, this process of compaction followed by dilation appears to begin and end at the same value of $\phi$ and for smaller $\gamma_r$ exhibits a $\gamma_r$ dependent strain value at which the peak $\phi$ is reached. This strain at peak $\phi$ is plotted against $\gamma_r$ in figure \ref{fig:DensityAndStress}, and a similar crossover from a $\gamma_r$ dependent regime near $\gamma_r$=0.15 is again observed.  

	A second observed bulk property that shows this crossover behavior is the shear stress $\sigma_{xy}$.  The rate of change of $\sigma_{xy}$ again shows a $\gamma_r$ dependence.  We illustrate this result by calculating the difference in $\sigma_{xy}$ between the last two average strain steps before reversal, which again shows a crossover from a $\gamma_r$ dependent regime near $\gamma_r$=0.15 (see figure \ref{fig:DensityAndStress}).  
 	We also note that averaged over many cycles, $\sigma_{xy}$ is quite hysteretic.  Very quickly the stress state of the system even at the beginning and end of each cycle is anisotropic, oscillating on average between two extremal values of $\sigma_{xy}$.
  
	Finally, we note that long time transients exist particularly for small $\gamma_r$.  The essential conclusion from these results is that a crossover in dynamics and bulk properties as a function of $\gamma_r$ occurs.  Our finding is insensitive to the transient: $\phi$ and $\sigma_{xy}$ increase for lower $\gamma_r$ over many cycles initially, but the strain at peak $\phi$ and slope of $\sigma_{xy}$ quickly become independent of cycle number.  As for the dynamics, the MSD curves shift to lower squared displacements, but the slopes do not change significantly.

{\it Connecting to Models of Subdiffusion ---} With such a strong correspondence between changes in bulk properties, and a crossover from a subdiffusive regime to simple diffusion, our next goal is to identify models that can capture the relevant subdiffusive dynamics.  A concise survey of stochastic processes that result in subdiffusive dynamics can be found in \cite{CodifiedKineticEquations}.  These models are all able to exhibit a subdiffusive regime described by an MSD which is $\propto t^{\beta}$, but are motivated by different physics.  In brief, there are two basic classes, fractional Brownian motion (fBm) and fractional time processes (ftp).  Both exhibit self-similarity under time dilation.  However in the case of fBm a time dependent diffusion coefficient is introduced, while for ftp a heavy-tailed distribution of waiting times is introduced.  As a result, the latter process is considered non-Markovian.  In addition, both processes can be generalized to fully fractional forms by introducing heavy-tailed displacement distributions in analogy to Levy flights.  The PDFs for all of these processes must obey a particular scaling form that reflects the self-affinity: 

\begin{equation}
t^{H} P\left(r,t\right)\propto\psi\left(\xi=\frac{r}{t^{H}}\right) 
\label{eq:scaling}
\end{equation}

\begin{figure}[h]
\centering
\subfigure{\includegraphics[width=0.40\textwidth]{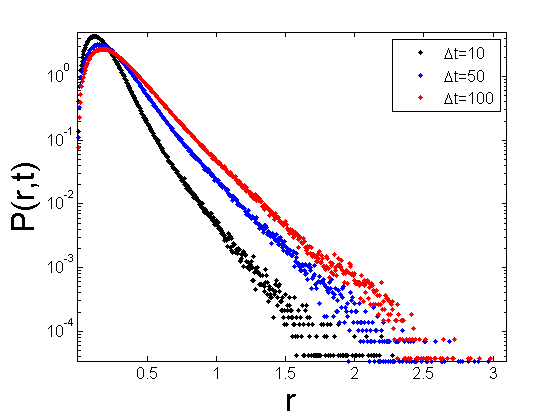}}\vspace{-.3cm}
\subfigure{\includegraphics[width=0.40\textwidth]{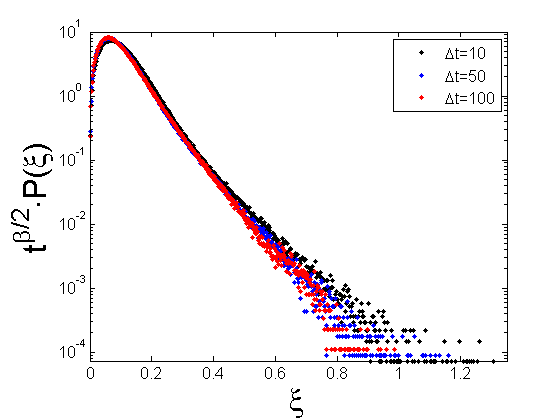}}
\vspace{-.3cm}
\caption{(Top) The unscaled displacement distributions for cycle intervals $\Delta t={10,50,100}$ (for black, blue and red).  (Bottom) the displacement distributions scale according to equation \ref{eq:scaling}, using the slope from the time averaged MSD for $\gamma_r=0.075.$}
\label{fig:Collapses}
\vspace{-.35cm}
\end{figure}

The function $\psi\left(\xi=r/t^{H}\right)$ will have different forms for the different models discussed.  Importantly, the self-similarity exponent $H=\beta/2$ in the cases of fBm and ftp, while the fully fractional processes do not generally satisfy these relationships.  Motivated by this fact, we scale PDFs for each of our trajectories at different $\gamma_r$ using the exponents gained from the time averaged MSDs.  As an example, Figure \ref{fig:Collapses} shows PDFs for the time intervals $\Delta t=10,50,100$ for the case of $\gamma_r=0.075$, scaled according to equation \ref{eq:scaling} with $H=\beta/2$.  Collapses for other $\gamma_r$ are of similar quality.  We conclude then that the fully fractional processes are not the appropriate model for the subdiffusive dynamics observed here, but that ftp and fBm are both viable candidates.

\begin{figure}[t]
\centering
\subfigure{\includegraphics[width=0.32\textwidth]{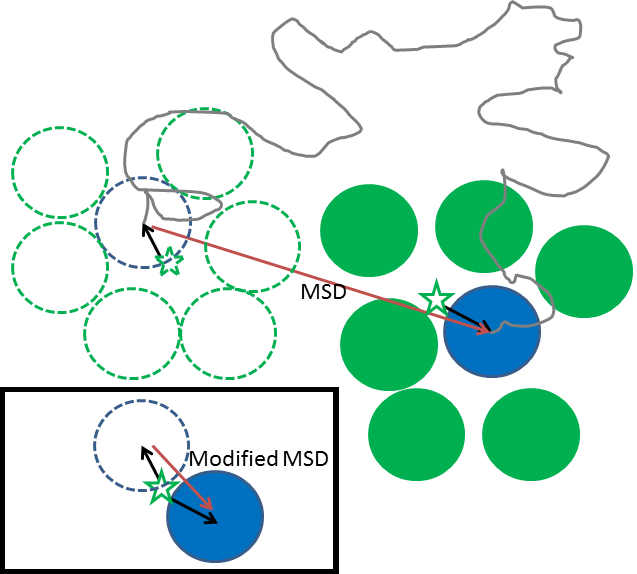}}
\vspace{-.35cm}
\caption{The above diagram illustrates the modification to the standard MSD approach.  In the standard approach, grain displacements are calculated relative to their own locations at a previous time (the displacement vector, in red, is labeled `MSD').  For the modified MSD (mMSD), described by the inset, particle displacements are calculated relative to the centroid (green star) of the nearby 6 grains (in green).  As the green grains move, the centroid moves as well, and the displacement of the blue grain is measured relative to this point.  The displacement vector associated with the mMSD is labeled in the bottom left.}
\label{fig:MSDDiagram}
\vspace{-.35cm}
\end{figure}	
{\it Modifying the MSD ---} To further differentiate between ftp and fBm, we consider the physical implications of waiting time distributions.  Waiting times can be modeled by continuous time random walks (CTRWs) in which the random walker has the ability to wait in one place rather than moving.  When the amount of time associated with waiting in one place has a heavy-tailed distribution, a ftp results.  This description shares similarities with the `cage breaking' scheme used to describe long relaxation times in granular and glassy systems.  In this scheme, particles such as the grains in our simulation `rattle' inside a cage of their neighbors on a shorter time scale, intermittently escaping out and into a new cage.  The time spent in cages is akin to a `waiting time.'  The ftp model would suggest that these waiting times should be heavy-tail distributed if the cage effect is to result in subdiffusive MSDs.  However, the notion of a cage is difficult to define in a system where all of the particles are in motion, so that the cage itself is a transient feature.  

\begin{figure}[t]
\centering
\subfigure{\includegraphics[width=0.4\textwidth]{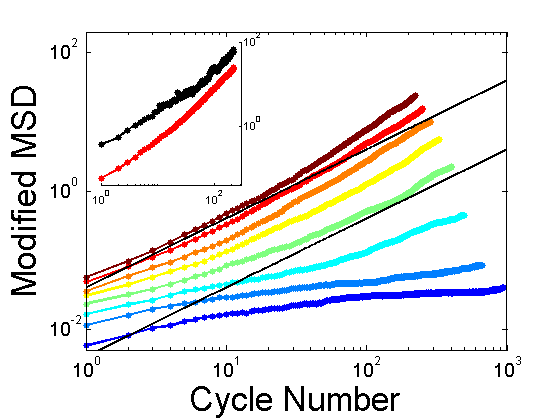}}
\vspace{-.35cm}
\caption{The mMSD, relative to the packing configuration at $t=0$, is shown.  Solid black lines have a slope of 1. (Inset) MSDs for $\gamma_r=0.225$ for MSD (black) and mMSD (red).}
\label{fig:ModifiedMSD}
\vspace{-.35cm}
\end{figure}

	To further explore the validity of this cage breaking scheme, we modify the MSDs (see figure \ref{fig:MSDDiagram}).	 To highlight motion relative to the cage, we calculate a modified MSD (mMSD) for a given grain $i$ with respect to a neighborhood of nearby grains.  In the isotropic state of the packing ($\gamma=0$ for the first cycle) the 6 nearby grains are found for grain $i$.  A centroid for these nearby grains is calculated at every strain step, and the displacement of grain $i$ relative to the centroid is calculated.  In this way, we characterize the motion of a grain relative to its nearby grains, which provides better insight into the nature of the local fluctuations in grain displacements.  Figure \ref{fig:ModifiedMSD} shows that the mMSD curves are very smooth, with total squared displacements almost an order of magnitude smaller than those found with the MSD.  These results suggest that the motions of the grains are coordinated with the motions of their neighborhoods of grains, which is itself changing size and shape over time.  Interestingly, unlike the MSD, the mMSD exhibits a continuous change from subdiffusive to superdiffusive motion, with superdiffusion for $\gamma_r \geq 0.15$.  Understanding the origin of superdiffusion in this context is a goal for future investigations.    

{\it Discussion ---} We have presented a systematic study of reversal amplitude $\gamma_r$ in a model amorphous system under cyclic shear.  Results for bulk and grain dynamical properties suggest a crossover in behaviors from subdiffusive to diffusive behavior near $\gamma_r=0.15$.  Interestingly the MSD exhibits a novel crossover in particle dynamics as a function of increasing $\gamma_r$, that is distinct from the crossover from `caged dynamics' to diffusion as a function of time reported in thermally driven systems such as colloids\cite{WeeksColloids}.  Temperature drives the system at the particle scale, while cyclic shear drives the system at the boundary, i.e. the driving force is applied at large scales.  
	
	Together these results indicate that driving from large scales rather than at the particle level fundamentally alters particle motion: the observed displacement distribution and its scaling behavior are not consistent with a simple cage-breaking model that is typically used for thermal systems.  Furthermore, a  mMSD measurement that characterizes grain motion relative to its neighbors, suggests that a `cage breaking' scheme may not provide the appropriate description of grain dynamics under cyclic shear: instead of rattling in their cage, grain MSDs are almost an order of magnitude smaller when measured relative to their cage.  

\begin{figure}[b]
\vspace{-.45cm}
\centering
\subfigure{\includegraphics[width=0.4\textwidth]{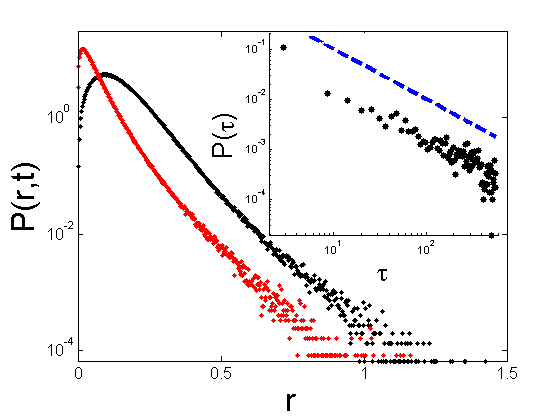}}
\vspace{-.35cm}
\caption{The single cycle displacement PDFs as a function of displaced distance $r$ is shown for $\gamma_r=0.075$, for both the displacements (black) and the relative displacements (red).  (Inset) Waiting times, as defined in the text, appear to have a heavy tailed distribution (again the example of $\gamma_r=0.075$ is shown).  Dashed line has a slope of -1.}
\label{fig:WaitingTimes}
\vspace{-.35cm}
\end{figure}

	To expand on this idea further, we look at displacement distributions relative to the neighborhoods.  The approach used here is to find the centroid of nearby grains during each cycle, and find the relative motion after one cycle.  For an example trajectory with $\gamma_r=0.075$, modified displacement distributions and standard displacement distributions for $\Delta t=1$ are shown in figure \ref{fig:WaitingTimes}.  All displacements are much smaller for the modified case, with a mean distance after one cycle approximately an order of magnitude smaller for the displacements relative to the neighborhoods.  If we consider the amount of time a grain spent inside its neighborhood (defined as being within the grain radius from the centroid) to be analogous to a cage lifetime, we find that the distribution of cage lifetimes appears to be scale invariant, with an apparent power-law distribution (inset of figure \ref{fig:WaitingTimes}).  This observation is consistent with the ftp model, where waiting times are assumed to be heavy tailed, although the exponent of the observed cage lifetimes (which appears to be 1) is not consistent with ftp.  Still, the observation that the cage lifetimes lacks a single characteristic time scale is at odds with the standard caging scenario, such as presented in \cite{WeeksColloids}, where the cages have a single characteristic lifetime for a particular density.  

Similar questions concerning the origins of subdiffusion in dense systems are ubiquitous in biological physics as well.  Subdiffusion has been observed in for instance, the cytoplasm of living cells \cite{EnhancedDiffCaspi,CrowdedFluidsWeiss} and mRNA dynamics in E. Coli \cite{GoldingCyto}.  Interestingly, a discussion has emerged as to whether ftp or fBm is most appropriate in the case of mRNA dynamics as well \cite{KlafterSimpleTest}.  

Beyond the implications for granular dynamics suggested by the ftp model, such as long-time correlations, the observation that the crossover in the slopes of the MSDs from less than 1 to 1 is consistent with the crossover in bulk density and stress properties has futher implications.  Since the origin of the collapse of the displacement PDFs is the self-affinity of these PDFs, and the self-affinity exponent controls the slope of the MSDs, we suggest that this self-affinity may be the origin of the crossover in all bulk properties as a function of $\gamma_r$, although the reason for such a strong dependence on the self-affinity exponent $H$ is not yet clear to us.  We suggest that a more complex picture of particle dynamics in cyclic, athermal, flows is needed, which will require more focus on the importance of global boundary forcing versus particle scale thermal forcing as well as driving amplitude.  

\begin{acknowledgments}
The computational work reported here was done largely on the University of Maryland High Performance Computing Cluster (HPCC).  This work was supported by NSF grant DMR0907146.
\end{acknowledgments}

\bibliography{Simsbib}

\end{document}